\documentclass[aps,prd,showpacs,nofootinbib,preprintnumbers,twocolumn]{revtex4}
\usepackage{bm}
\usepackage{latexsym}
\usepackage{natbib}
\usepackage{url}
\usepackage{dcolumn}
\usepackage{color}
\usepackage{amsfonts,amssymb,amsmath}
\usepackage{graphicx,epsfig}
\usepackage{psfrag}
\usepackage{subfigure}
\usepackage{natbib}
\def\pbest{\texttt {Pbest~}}
\def\gbest{\texttt {Gbest~}}

\def\As{$ 10^{9} A_s$}

\begin{document}
\title{Revisiting  Cosmological parameter estimation}
\author{Jayanti Prasad}
\email{jayanti@iucaa.ernet.in}
\affiliation{IUCAA, Post Bag 4, Ganeshkhind, Pune 411007, India.}
\date{\today}
\begin{abstract}
Constraining theoretical models with measuring the parameters of those from 
cosmic microwave background (CMB) anisotropy data is one of the most
active areas in cosmology. WMAP, Planck and other recent experiments 
have shown that the six parameters standard $\Lambda$CDM cosmological model  still  best fits
the data. Bayesian methods based on Markov-Chain Monte Carlo (MCMC) sampling  
have been playing leading role in parameter estimation from CMB data. 
In one of the recent studies \cite{2012PhRvD..85l3008P}  
we have shown that particle swarm optimization (PSO) 
which is a population based search procedure can also be effectively  used to  
find the cosmological parameters which are best fit to the WMAP seven year data. 
In the present work we show that PSO not only can find the best-fit point,
it can also sample the parameter space quite effectively, to the extent that we can 
use the same analysis pipeline to process PSO sampled points which is 
used to process the points sampled by Markov Chains, and get consistent results. 
We also present implementations   of downhill-simplex Method of Nelder and Mead
and Powell's method of  Bound Optimization BY Quadratic 
Approximation (BOBYQA) in this work for cosmological parameter estimation, 
and compare these methods with PSO. Since PSO has the advantage that it only needs the search 
range and does not need covariance-matrix, starting point or any other quantity which depend 
on the final results, it can be quite useful for a blind search of the best fit parameters. 
Apart from that,  PSO is based on a completely different algorithm  so it can  
supplement MCMC methods. We use PSO to estimate parameters from the WMAP nine year and 
Planck data and get consistent results.
\end{abstract}
\maketitle
\section{Introduction}
Cosmological parameters characterizing  the primordial curvature and tensor 
perturbations which were present at the end of inflation, background expansion
of the universe, physical events (like reionization) that took place after 
the decoupling of CMB photons can be estimated from the temperature and 
polarization anisotropies present in the CMB sky 
\cite{1994PhRvL..72...13B,1995PhRvD..51.2599H,1995ApJ...439..503D,1996PhRvD..54.1332J,1996PhRvL..76.1007J,1997ApJ...488....1Z,1997MNRAS.291L..33B,1998PhRvD..57.2117B, 2002PhRvD..66f3007K,2003ApJ...596..725C,2003moco.book.....D}.
CMB anisotropies are represented by a stochastic field on a 2-sphere
and can be completely characterized by the correlations between different directions.
If CMB anisotropies are statistically isotropic and Gaussian also, as most observations indicate,
then they can be completely  described by two-point correlations 
or the angular power spectrum 
\cite{1987MNRAS.226..655B,1995PhRvD..51.2599H,1997MNRAS.291L..33B,1998PhRvD..57.2117B,2001CQGra..18.2677C,2004astro.ph..3344C,2008PhRvD..77j3013H,2009PhRvD..79h3012H}. 
Since CMB anisotropies are quite small (one in one hundred thousand part for temperature)
their evolution can be studied quite effectively using the theory of linear perturbations 
\cite{1984ApJ...285L..45B,1995PhRvD..51.2599H,1995ApJ...455....7M,1998PhRvD..57.3290H}.
There are numerical codes available which give us the angular power spectrum 
of CMB anisotropies at present for a given model of primordial fluctuations and background 
cosmology \cite{1996ApJ...469..437S,1999ascl.soft09004S,2000ApJ...538..473L} 
and that can be  compared  with the observed angular power spectrum we get from CMB 
experiments like WMAP and Planck.  

In general cosmological parameter estimation from CMB data involves finding the point in 
the multi-dimensional parameters space at which the likelihood function 
is maximum i.e., the best fit point.   In Bayesian framework this is done by 
sampling the joint posterior probability distribution  and from that 
various statistical quantities are computed. 
Since grid based sampling is  prohibitively expansive (computationally) for a large number 
of parameters (at least six in our case), stochastic methods are generally employed which scales 
linearly (at the most) with the number of parameters. Stochastic methods based on Markov-Chain 
Monte Carlo (MCMC) sampling employing Metropolis-Hasting algorithm \cite{1953JChPh..21.1087M,Hasting1970}
have been leading  cosmological parameters estimation from CMB data. MCMC method  
have an interesting property that they ensures that the number density of the sampled points 
is asymptotically  proportional to the joint probability distribution 
\cite{2001CQGra..18.2677C,2002PhRvD..66j3511L,2003ApJS..148..195V,2010LNP...800..147V,2014JCAP...07..018D}. 
There has been presented some modification in the usual Metropolis-Hasting algorithm to 
achieve better performance \cite{2014JCAP...07..018D}. 
However, all the methods based on MCMC share the common weakness that the step size has to 
be chosen very carefully, in particular when the likelihood surface has multiple local maxima. 

In a recent study  \cite{2012PhRvD..85l3008P} we have shown that particle swarm 
optimization or PSO (we call our code {\tt COSMOPSO}),  which is an artificial intelligence inspired 
population based search procedure, can be used to estimate cosmological parameters from the WMAP seven year data. 
At that time it was not clear to us whether PSO can be used as a ``sampler'' also like MCMC,
apart from an optimizer. By carrying out a large number of simulations, 
we have found that if the convergence of PSO can be delayed and we use multiple realizations of
PSO then PSO can also sample the parameter space quite  effectively to the extent that we 
can use the same analysis pipeline to process the PSO sampled points which is used to process the
points sampled in MCMC i.e., {\tt GetDist}. 


The plan of this paper is as follows. In section (2) we show that PSO particles in {\tt COSMOPSO} 
do sample the parameter space effectively and present the results which we get for the 
WMAP nine year and Planck data. In section (3) we present an implementation of
the downhill simplex method of Nelder and Mead \cite{neldermean1965}  for cosmological
parameter estimation (we call the code {\tt COSMODSM}) for the WMAP nine  year data. 
In section (4) we present an implementation of Powell's method of Bound Optimization BY Quadratic 
Approximation (BOBYQA)  for cosmological parameter estimation (we call our code {\tt COSMOBOBYQA})
and  compare {\tt COSMOMC}, {\tt COSMOPSO}, {\tt COSMODSM} with {\tt COSMOBOBYQA} and present 
results for WMAP nine year data. In section (5) we present the discussion and conclusions of our work.
\section{Particle Swarm Optimization}
\begin{figure*}[t!] 
\begin{center}
\epsfig{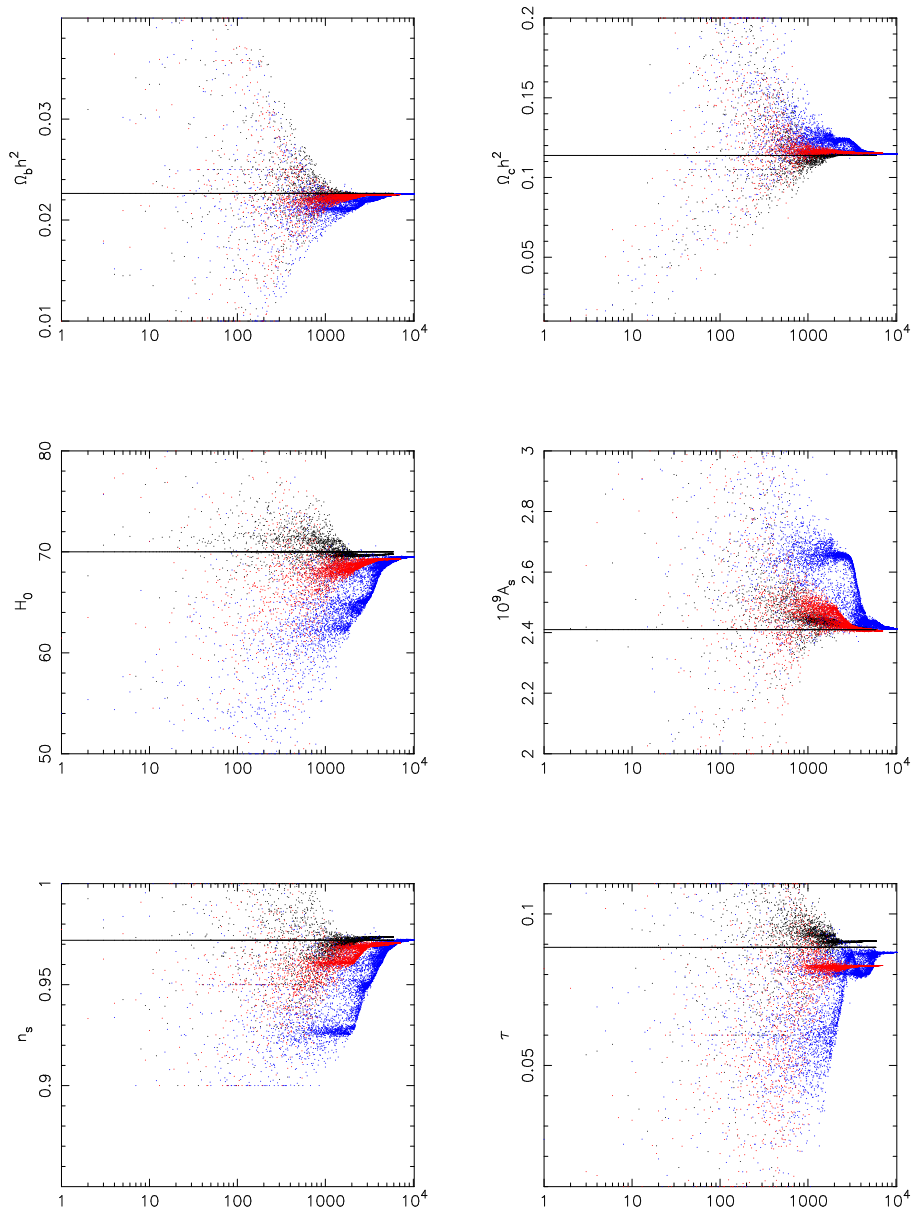}
\caption{The figure shows the positions of PSO particles along different directions 
with iterations (x-axis) for three different realizations (shown by different colors).
PSO particles not only finally reach the point at which the cost function (likelihood) is maximum
they also make random walks in way that the size of the jump at every stage is 
roughly  proportional to their distance from the global maximum. As a  result of this 
type of random walk the density of sample points increases when we approach towards the global 
maxima.
In the above figure the distribution of particles along x-axis is not important
since the number of steps which are needed for convergence are different for different
realizations.}
\label{fig:pso_samples}
\end{center}
\end{figure*}
PSO was proposed by Kennedy and Eberhart \cite{KennedEberhart1995,KennedEberhart2001} 
in 1995 and since then there have been proposed many modifications of PSO for different type
of applications (for recent updates see \cite{Blum2008,Lazinic2009}). 
PSO is a population based search procedure which carry out local and global search in a
multi-dimensional parameter space simultaneously \cite{KennedEberhart1995,KennedEberhart2001,Engelbrech2002,Blum2008,Lazinic2009}.
PSO has been regularly  used in engineering problems since its inception, however, recently it has become 
common in astronomy also \cite{2005MNRAS.359..251S,2010PhRvD..81f3002W,2011ApJ...727...80R,2012PhRvD..85l3008P,2013PhRvD..87j4021L}. 
In PSO, a team of particles or computational agents is launched in the multi-dimensional parameter space which is driven by the 
exploration of all the individual particles and the team together.
If the position and the velocity of $i^{th}$ PSO particle at ``time'' $t$ are represented by $X^i(t)$
and $V^i(t)$  respectively then according to the PSO algorithm the velocity $V^i(t+1)$ at time $(t+1)$ is
computed as 
\begin{eqnarray}
V^i(t+1) = w V^i(t) + c_1 \xi_1 [P^i-X^i(t)]  \nonumber \\ 
+  c_2 \xi_2 [ G- X^i(t)],
\label{eqn:pso1}
\end{eqnarray}
and the position is updated as:
\begin{equation}
X^i(t+1) = X^i(t) + V^i(t+1). 
\label{eqn:pso2}
\end{equation}

In Eq.~(\ref{eqn:pso1})  $P^i$ is the point at which the $i^{th}$ particle has found the 
maximum value of the cost function $f(X) $ (in our case minimum value of $- \log {\mathcal L}$ 
where ${\mathcal L}$ is the likelihood function)  and is called its  \pbest or the 
``personal best' and $G$ is the location of that \pbest which is highest and called 
the ``global best'' or \gbest. The coefficients $c_1$ and $c_2$ are called acceleration 
coefficients and their values determine the weight we want to give to the exploration of 
the individual particles and the team  respectively.
The coefficient $w$ is called the inertia weight and its value decides the weight which we 
want to give the ``inertial'' motions of the particles and $\xi_1$ and $\xi_2$ are two uniform random  
numbers in the range $[0,1]$.

The second and the third term in the right hand side of Eq.~(\ref{eqn:pso1}) resemble a  simple harmonic 
motion (Hooke's law)  and corresponds to the force $F^i(t)$ acting on particle $i$ at time $t$:
\begin{equation}
F^i(t)  = -k  [X^i(t) -{\tilde X}],
\label{eqn:pso_shm} 
\end{equation}
where ${\tilde X}$ is the location of the \pbest or \gbest and $k=c_i\xi_i,i=1,2$, is the 
force constant which is a stochastic variable here. Eq. ~(\ref{eqn:pso_shm}) shows that PSO
particle oscillate around the \pbest and \gbest with gradually decreasing  random amplitudes. 
In the beginning of PSO particles exploration there are a very few particles which have 
their \pbest close to the \gbest (one particle has its \pbest and \gbest the same), however,
when the particles approach close to the global maximum of the cost function all the 
particles have their \pbest close to the \gbest and they all oscillates around the 
points which are very close to each other. 

In order to show the behavior of the cost function $f(X)$ close to the best-fit point
we  Taylor expand   $f(X)$ around the best fit point $X_0$, in the following way:
\begin{equation}
f(X) = f(X_0) + \frac{\partial^2 f(X)}{\partial X^2}|_{X=X_0} (X-X_0^2) + ...,
\end{equation}
which shows that very close to the best fit point $X_0$, every cost  function $f(X)$ can be approximated as 
a quadratic function so it is not required that the sampling method we use have more
complex strategy than what is needed to sample a quadratic function. Since step 
size of PSO particles is approximately proportional to their distance from the global maximum 
(the approximation becomes better when we reach close to the global maximum) therefore
the density of sample points increases in a natural way.  
We support the claim that PSO also can sample the parameter space effectively by estimating 
the cosmological  parameter estimation from  the WMAP nine year and Planck data. 
We  arrange the points which are sampled by PSO into ``chains''  and use the {\tt\ COSMOMC} analysis 
pipeline {\tt GetDist} on that.  

In any PSO implementation we not only need to chose the values of the design 
parameters $c_1,c_2, w$ and the number of PSO particles, we also need to
select a method for setting the initial positions and velocities of particles,
maximum velocity, boundary condition and a termination criteria. Apart from 
the termination criteria, we keep the values of the design parameters and
other consideration the same as we used in  \cite{2012PhRvD..85l3008P}. 
For termination, we stop PSO exploration when find that the change in the 
cost function is smaller than a user given value for a given number of steps. 

Our cosmological parameter estimation  code based on particle 
swarm optimization named  {\tt COSMOPSO} has three main components
which are used for computing theoretical $C_l$s, computing likelihood and 
evolving PSO particles.   Since we have already discussed the 
third component,  here we would like summarize the first and the second
components. 

For computing the angular power spectra of  CMB temperature 
and polarization anisotropies for a given theoretical model (set of cosmological 
parameters)  we use publicly available code {\tt CAMB} \cite{camb} which employs
a line of sight integration approach as was given in  \cite{1996ApJ...469..437S}. 
We use April 2014 version of {\tt CAMB} which is different than what was
used in WMAP nine year and Planck data analysis so some difference are 
expected in the values of cosmological parameters we report here. 

In {\tt CAMB} we vary only six cosmological parameters named physical densities of baryons ($\Omega_bh^2$) 
and dark matter ($\Omega_ch^2$), the Hubble parameter at present ($H_0$), the amplitude ($A_s$) and 
spectral index ($n_s$) of primordial scalar perturbations  at some pivot scale ($k_{*}$), 
and the the optical depth of reionization ($\tau$) epoch.  
Note that we consider $H_0$ as a fitting parameter and not the angular 
size ($\theta$) of the last scattering surface as is commonly done (in {\tt COSMOMC}) and have some advantages over $H_0$.  
We do not consider dark energy density ($\Omega_{\Lambda}$) as a fitting parameter and compute that 
from the  flatness condition:
\begin{equation}
\Omega_{\Lambda} = 1 -\frac{\Omega_bh^2+\Omega_ch^2}{(H_0/100)^2}.
\end{equation}

We use pivot scale for the scalar and tensor power spectrum $0.002\text{Mpc}^{-1}$, 
when using only WMAP nine year data and use pivot scale $0.05\text{Mpc}^{-1}$, when 
using WMAP nine year and Planck data combined. 
We use {\tt l\_max\_scalar=2600, l\_max\_tensor=1500, l\_eta\_max\_scalar=4000, l\_eta\_max\_tensor=3000},
the width of reionization redshift $\Delta z =0.5$, CMB temperature $T_{cmb}=2.72548$, Helium 
fraction $Y_{He}=0.24$, massless neutrino species $n_{\nu}=3.04$. We keep  logical variables {\tt AccurateReionization} 
and {\tt AccurateBB} true and do not compute tensor perturbations and keep {\tt WantTensors} false in {\tt CAMB}.

The procedure which we follow to estimate cosmological parameters is as follows. 
We use eight different realizations (different seeds for random number
generator) of PSO which have been converged  for sampling. For every realization we sorted the 
sample points on the basis of the likelihood value and wrote them in the form of chains as are 
needed by the {\tt GetDist} program of {\tt COSMOMC} to compute the best-fit values, 
68\% limits and confidence regions. In order to make one-dimensional probability distribution
(marginalized)  and two-dimensional contours (68\% and 95\% confidence regions)  smooth,
smoothing parameters are passed to {\tt GetDist}  parameter file.  We found that PSO needs
more smoothing than what is needed for MCMC samples. 
\subsection{WMAP nine year data}
\begin{table*}[t!]
\small
\begin{center}
\begin{tabular}{cccl}  \hline \hline
S. No & Parameter & Prior & Description \\ \hline 
1     & $\Omega_b h^2$  & $[0.01,0.04]$ &  Baryon density today \\
2     & $\Omega_c h^2$  & $[0.01,0.20]$ & Cold dark matter density today \\
3     & $H_0$  & $[50,80.0]$ &  Hubble parameter at present \\
4     & \As    & $[2.0,4.0]$ & Primordial curvature perturbations  \\
5     & $n_s$ & $[0.9,1.0]$  & Scalar spectrum power-law index   \\
6     & $\tau$& $[0.01,0.11]$ & Thomson scattering optical depth due to reionization \\ \hline  
\end{tabular}
\caption{Cosmological parameters and their priors we used in {\tt COSMOPSO}, {\tt COSMODSM} and {\tt COSMOBOBYQA}.  
Note that the range we have considered is different than what is generally used (in 
{\tt COSMOMC}) mainly because all the three methods we considered demand that we must be 
able to compute the likelihood function at every point in the prior range.} 
\label{tab:prior}
\end{center}
\end{table*}
In order to estimate cosmological parameter with {\tt COSMOPSO} and demonstrate that 
PSO can also  sample the parameter space quite effectively, apart from finding the best 
fit values, here we present  the result for the WMAP nine year data 
which has been made publicly available by the WMAP team \cite{wmap}. 
WMAP team has also made a FORTARN code available with the data  to compute likelihoods for
CMB temperature and polarization data for a user given theoretical model i.e., theoretical $C_l$s.  
There have been a few changes in the way likelihoods are computed for the nine year 
release \cite{2003ApJS..148..195V}, however, the basic approach has been the same which was outlined 
in \cite{2003ApJS..148..175S}.  
WMAP nine year likelihood code computes likelihood differently at low and high-l for temperature
and polarization power spectra \cite{2013ApJS..208...20B}. At high-l ($ l > 32$) the likelihood is no 
longer computed  using {\tt MASTER} \cite{2002ApJ...567....2H} (as was the case for earlier data releases) 
and use  an optimally estimated power spectrum and errors based on the quadratic estimator
\cite{1997PhRvD..55.5895T}.  At low-l ($ l < 32$) the TT-likelihood is computed 
using the Blackwell-Rao estimator which employs Gibbs sampling and uses 
bias-corrected Internal Linear Combination (ILC) map as an input \cite{2004ApJS..155..227E}. 

WMAP nine year likelihood does not have any extra parameter and needs only the
theoretical power spectra for computing the likelihood. Some of the parameters 
which are used in the WMAP likelihood code are  {\tt l\_max=1200} and {\tt 800} and 
for TT and TE respectively. We also set {\tt use\_TT, use\_TE, use\_lowl\_TT, use\_lowl\_pol} 
true in the WMAP nine year likelihood code. 

\begin{figure*}[t!] 
\begin{center}
\epsfig{width=14cm,file=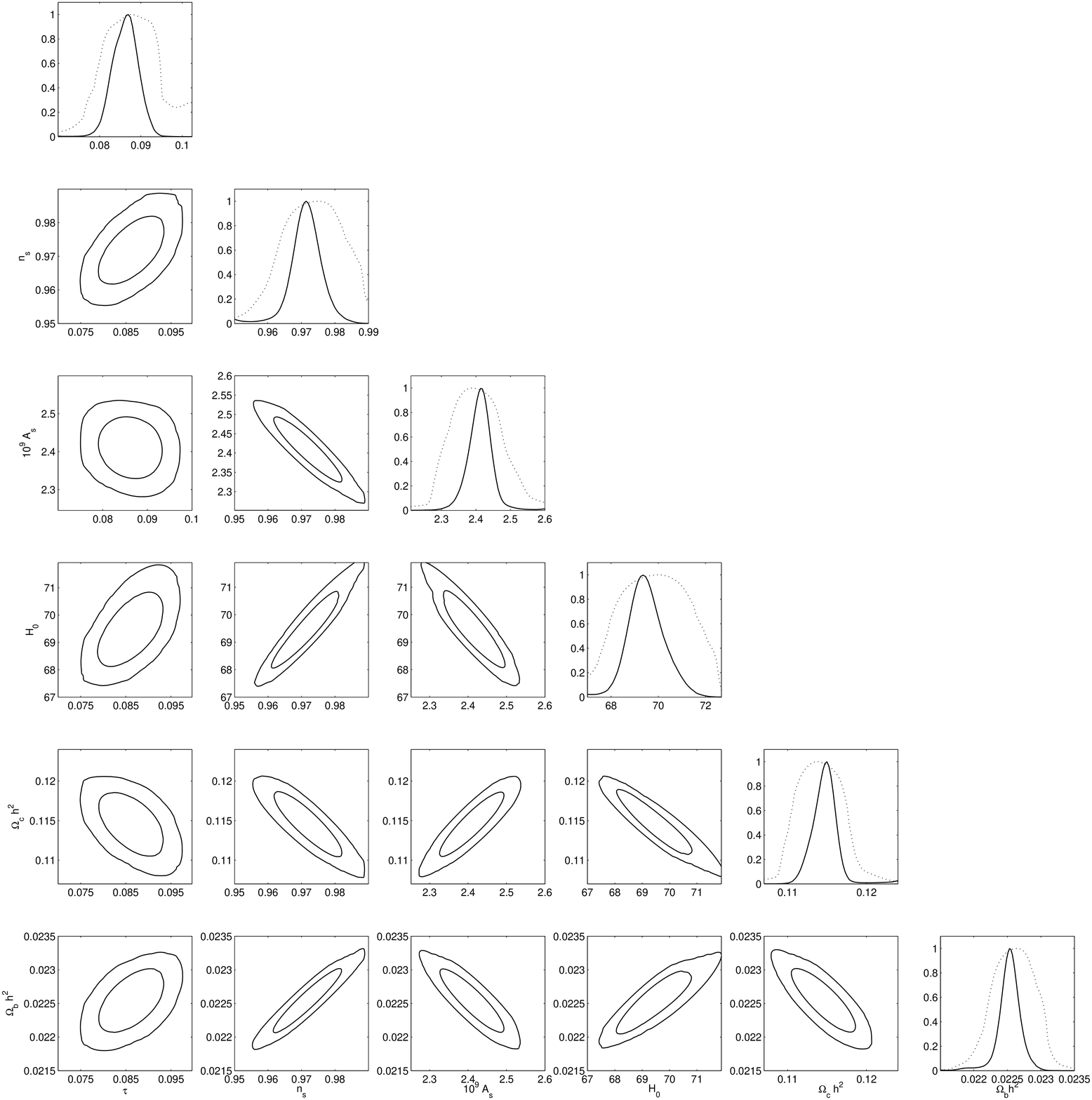}
\caption{The diagonal panels in this figure show the one-dimensional marginalized distributions
and other panels show two-dimensional marginalized constraints (68\% and 95\% CL) for  the
standard six cosmological parameters for the WMAP nine year data which we obtained 
by combining the sampled points  of eight different realizations of PSO and 
using the {\tt Getdist} (the program which is used for  analysis in {\tt COSMOMC})  on those. }
\label{fig:pso_cont_w9}
\end{center}
\end{figure*}

The prior which we use for cosmological parameters are given in Tab.~(\ref{tab:prior}).  Note that range 
which we consider here are  different than what are considered in {\tt COSMOMC} mainly because all the three 
methods which we consider in the present work demand that we should be able to compute the likelihood
function at all the points in the search space. 

\begin{table*}
\centering
\begin{tabular}{ccccc}  \hline  \hline 
\multicolumn{1}{c}{} & \multicolumn{2}{c}{PSO} &\multicolumn{2}{c}{COSMOMC}                 \\  \hline
Parameter        &  Best Fit  & 68\% Limit               & Best Fit   &  68 \% Limit           \\ \hline 
$\Omega_bh^2$    & 0.02260   & 0.02252 $\pm$  0.00018   &  0.02259   &  0.02265 $\pm$ 0.00048 \\ \hline 
$\Omega_ch^2$    & 0.1140     & 0.1148  $\pm$  0.0018    &  0.1122    &  0.1137  $\pm$ 0.0046  \\ \hline
$H_0$            & 69.827     & 69.397  $\pm$  0.9758    & 70.059     &  69.660  $\pm$ 2.178  \\ \hline
\As              & 2.395      & 2.416   $\pm$  0.044      &  2.369    &  2.402   $\pm$ 0.068   \\ \hline
$n_s$            & 0.9733     & 0.971   $\pm$  0.005     &  0.978     &  0.9730  $\pm$ 0.013  \\ \hline
$\tau$           & 0.086      & 0.086   $\pm$  0.003     &  0.086     &  0.088   $\pm$ 0.014  \\ \hline
-$\log {\mathcal L}$        & 3778.7840  &                          & 3779.0220  &                       \\ \hline  
\end{tabular}
\caption{Cosmological parameters estimated using {\tt COSMOMC} and PSO for the WMAP nine year temperature and 
polarization data. Note that the latest version of {\tt COSMOMC} computes $A_s$ at pivot scale $k_0$, 
$0.05~\text{Mpc}^{-1}$  so we convert the best fit and mean values to $0.002~\text{Mpc}^{-1}$ for comparison. }
\label{tab:pso_1}
\end{table*}

One of the results which we would like to highlight is that the sampling of the 
parameter space done by PSO particles. In Fig.~(\ref{fig:pso_samples}) we show 
the distribution of particles in PSO  along the directions of different cosmological 
parameters for three different realizations. From the distribution of particles we can see
that in a single realization most likely we will not have a symmetric
distribution of particles around the global best position.  However, when we use
multiple realizations of PSO which have been converged we get more and more
symmetric distribution.  Which is important if we want to sample the 
parameter space close to the global maximum fairly.
The final confirmation of the distribution of particles can be done by looking at the
one dimensional marginal distribution of cosmological parameters in 
Figs.~(\ref{fig:pso_cont_w9}) and  (\ref{fig:pso_cont_w9p}).

We show the best fit cosmological parameters which we estimate using {\tt  COSMOMC}
and {\tt COSMOPSO} for the WMAP nine year data with their 68\% limits in 
Tab.~(\ref{tab:pso_1}). Note that there is a slight difference between the 
{\tt COSMOMC} best-fit parameters which we present in Tab.~(\ref{tab:pso_1}) and
published in  \cite{2013ApJS..208...19H} which can be attributed to
the different versions  of codes we are using.  From the table we can see that 
PSO is not only able to find the  best-fit point,  it is also able to find the 
mean and 65 \% limits of the parameters quite close to what we get from {\tt COSMOMC}
for the same data. We would like to mention that the sampling done in {\tt COSMOPSO} is not 
as good as in {\tt COSMOMC} and as a results that we get slightly non-Gaussian 
one-dimensional marginal probability distributions as compared to {\tt COSMOMC}. 
We also find that the minimum value of $-\log {\mathcal L}$ which we get 
in {\tt COSMOPSO} is very close to what we get in {\tt COSMOMC} (in fact {\tt COSMOPSO} 
gives us  slightly better likelihood).

In Fig.~(\ref{fig:pso_cont_w9}) we show the one dimensional marginal probability 
distributions for the six cosmological parameters in the diagonal panels and 68\% 
and 98\% confidence regions in the rest of the panels. 
From these panels we see that the marginal probability distributions and 
two-dimensional joint probability distribution as shown by the contour plots 
show expected behavior. This clearly shows that the way PSO particles 
make random walks in the parameter space can be used for sampling the 
parameter space quite effectively. 

\subsection{WMAP nine year + Planck data}
\begin{figure*}[t!] 
\begin{center}
\epsfig{width=14cm,file=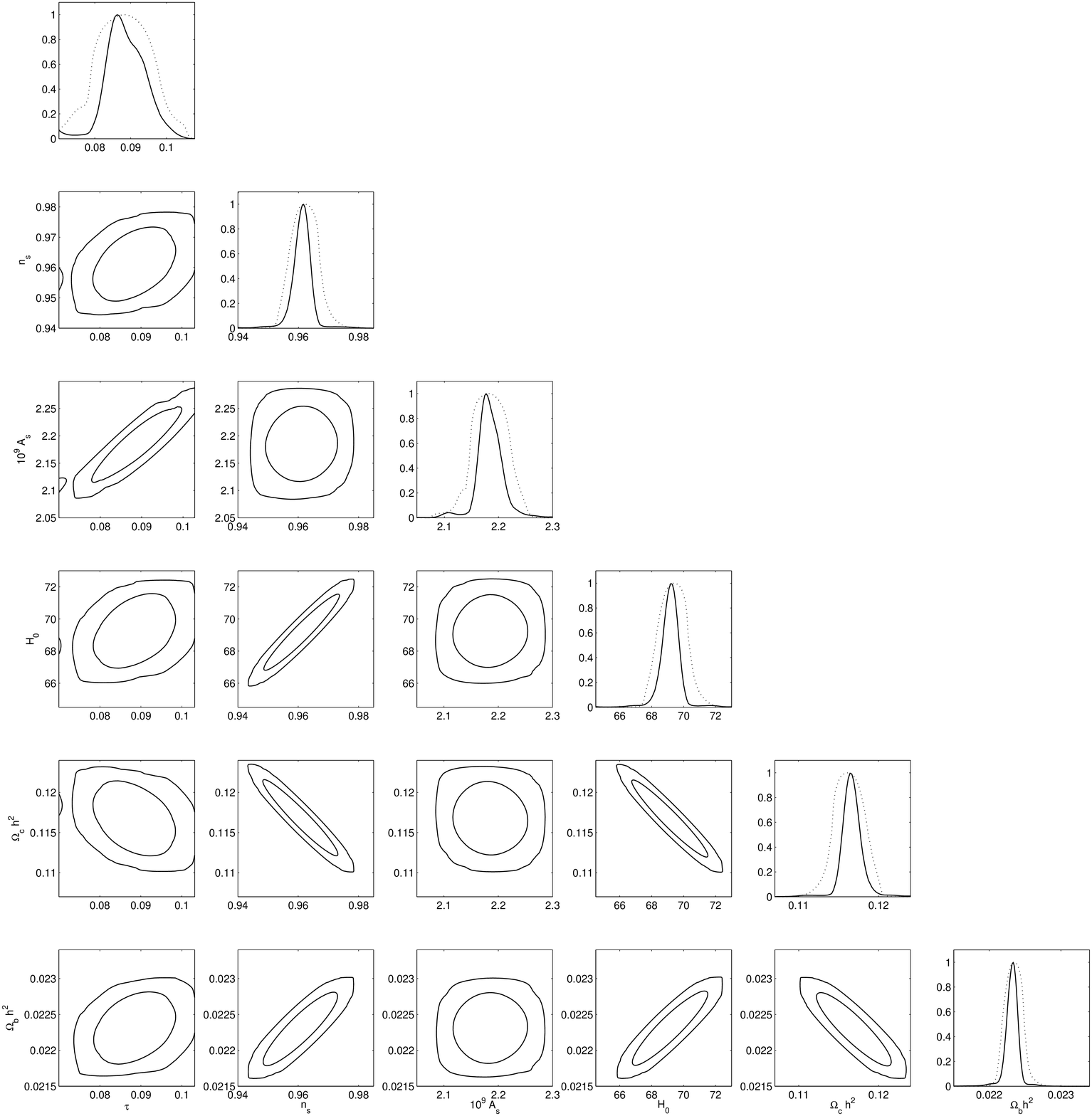}
\caption{The same as in Fig.~(\ref{fig:pso_cont_w9}) for the WMAP nine year + Planck data. }
\label{fig:pso_cont_w9p}
\end{center}
\end{figure*}

\begin{table*}
\centering
\begin{tabular}{ccccc}  \hline  \hline 
\multicolumn{1}{c}{} & \multicolumn{2}{c}{PSO} &\multicolumn{2}{c}{COSMOMC}                 \\  \hline
Parameter        &  Best Fit  & 68\% Limit                 & Best Fit   &  68 \% Limit           \\ \hline 
$\Omega_bh^2$    &  0.02237   & 0.02231   $\pm$ 0.00001    &  0.02233   & 0.02237  $\pm$  0.00020 \\ \hline 
$\Omega_ch^2$    &  0.1159    & 0.1167    $\pm$ 0.0014     &  0.1166    & 0.1163  $\pm$  0.0020 \\ \hline
$H_0$            &  69.508    &  69.138   $\pm$ 0.683      &  68.688    & 68.835  $\pm$  0.975 \\ \hline
\As              & 2.1759     &  2.1838   $\pm$ 0.023      &  2.1741    & 2.1832  $\pm$  0.037\\ \hline
$n_s$            &  0.9628    & 0.9610    $\pm$ 0.003      &  0.9614    & 0.9618  $\pm$  0.005 \\ \hline
$\tau$           &  0.087    &  0.088     $\pm$ 0.005      & 0.086      & 0.088  $\pm$  0.009 \\ \hline
-$\log {\mathcal L}$   & 8694.4260  &                      &  8693.6840 &                       \\ \hline  
\end{tabular}
\caption{Cosmological parameters estimated using {\tt COSMOMC} and PSO for the WMAP nine year (temperature and 
polarization data) + Planck (temperature data).}
\label{tab:pso_2}
\end{table*}

In this section we present our analysis for the WMAP nine year + Planck data.
A detail description of the Planck project and the data product is given in
\cite{2014A&A...571A...1P}, however, here we would like to mention that
Planck has more frequency channels over wide frequency range, higher angular
resolution and better noise sensitivity which makes it far better equipped to 
handle different type of systematic (foreground etc.) than WMAP. 
At present Planck has made only temperature data publicly available which can be used
alone or with a combination of other CMB data sets to constrain theoretical models.
Planck collaboration has also made a code named {\tt Plc} publicly available for
computing the  likelihood for temperature, polarization and lensing data.
The code can be downloaded from \cite{plc} and the detail explanation abut the
methodology of power spectrum and likelihood estimation 
can be found in \cite{2014A&A...571A..15P}. 

In order to compute likelihood for a given cosmological model, which is
represented by a set of six angular power spectra (TT, EE, BB, TE, TB, EB), 
we use Planck likelihood software {\tt Plc}.  
In Planck likelihood the high-l (up to  $l=2500$) TT likelihood is computed using 
{\tt CAMspec} which needs 14 extra (nuisance)  parameters which we  
fix to their best fit values given in \cite{2014A&A...571A..16P} 
and vary only the six cosmological parameters which directly affect
$C_l$s. Planck likelihood software computes low-l ($l=0$ to $l=49$) 
TT likelihood using {\tt commander} and low-l polarization likelihood
is computed using WMAP nine year data which needs TT, EE, BB and TE 
power spectra from $l=0$ to $l=32$. Since power at $l=0$ and $l=1$ does
not have any sense so we keep their values  $-1$ in the Planck likelihood. 
In {\tt COSMOPSO} we add all the three Planck likelihoods (after changing
the sign) to the WMAP nine year likelihood for our optimization function
or the cost function. In order to cross check that we get consistent  
results we compare those with {\tt COSMOMC} results. 
 
The best fit cosmological parameters which we get for the WMAP nine + Planck 
data which we get for {\tt COSMOMC} and {\tt COSMOPSO} are given in  Tab.~(\ref{tab:pso_2}).
From the table we can re-confirm that PSO not only can find the best
fit cosmological parameters,  which gives the maximum value of the likelihood function,
it can also find consistent estimates  of the 68\% limits of the cosmological parameters.
For the WMAP nine year + Planck data also we have consider 
eight different realization of {\tt COSMOPSO} which have been converged 
and create chains out of the sample points  to process with {\tt GetDist}. 
In Fig.~(\ref{fig:pso_cont_w9p}) we show one dimensional marginalized probability
distribution and 68\% and 95\% confidence regions of the estimated parameters 
and for those we can see that PSO gives consistent result for the joint data also.   

\section{Downhill Simplex Method (DSM)}
\label{ss:amb}
Downhill simplex method was introduced by Nelder and Mead \cite{neldermean1965} in 1965
exactly for the purpose we are using it here i.e., ``maximization of a likelihood function,
in which the unknown parameters enter non-linearly''.
On the basis of Nelder and Mead's algorithm \cite{1992nrfa.book.....P} has written 
a quite efficient program named `amoeba''. A  great detail about the 
algorithm  and implementation is also given in \cite{1992nrfa.book.....P}. 
In place of directly using amoeba we have written our code from the scratch  
on the basis of the algorithm given in the original paper \cite{neldermean1965}. 
Since there are some minor difference between the algorithm given in \cite{neldermean1965} and
implemented in \cite{1992nrfa.book.....P}, therefor we would like to give an overview of the 
algorithm which is quite close to what is given \cite{fh2010_optimization} also.

\subsection{Basics}
A geometric object  with $(n+1)$ vertices in a n-dimensional space is called a ``simplex''.
For example, a triangle in a two-dimensional flat surface is a simplex. 
In one-dimensional optimization (root finding for example) we can ``bracket''
the solution between two points and keep making our bracket smaller and smaller
following some algorithm and finally reach close to the solution. 
In more than one dimensional space there are no ``bracketing'' methods available
and so the problem is much more challenging. 

In the downhill simplex method we begin with an initial simplex i.e., set of
$(n+1)$ points in a $n$ dimensional space and replace the point at which 
the cost function is maximum (the worst point)  with a new point which we 
obtain by carrying out three type of geometrical operations named reflection, 
expansion and contraction.  We keep repeating the process until the values of the 
function at all the $(n+1)$ points do not become close to each other (within a user defined 
tolerance).  In downhill simplex method the simplex adapts the local geometry, 
 in the sense,  that it gets elongated when encounter a long inclined place, reflect when 
face a valley at an angle and contract close to a minimum. 

The most important consideration of downhill simplex method are (1) the choice of the initial 
simplex (2) tolerance and (3) the values of the expansion and contraction coefficients. 
We keep the values of reflection coefficients $\alpha$,  expansion coefficient $\beta$ and the 
contraction coefficients $\gamma$ to  their standard values $-1$ and $2$ and $1/2$ respectively. 
Choosing  tolerance is easy (we can chose a sufficiently small number) but choosing an initial simplex 
is a bit harder since we want to make sure that the simplex always remain within the allowed range
without imposing any boundary condition.  
\subsection{The Algorithm}

\begin{table*}
\begin{center}
\begin{tabular}{cccccccccc} \hline  \hline 
S. No &  (a,b) & iterations & $\Omega_bh^2$ & $\Omega_ch^2 $ & $H_0$ & $A_s$ & $n_s$ &  $\tau$ & $-\log {\mathcal L}$    \\ \hline 
1 & (0.20,0.60)& 428    & 0.022546 & 0.113998 & 69.689714 & 2.400900 & 0.971840 &  0.085522 &  3778.8044 \\ \hline 
2 & (0.20,0.40)& 330   & 0.022648 & 0.114334 & 69.782342 & 2.395106 & 0.974042 &  0.086739 &  3778.7934 \\ \hline 
3 & (0.25,0.40)& 348   & 0.022603 & 0.114112 & 69.811296 & 2.396340 & 0.973377 &  0.086554 &  3778.7840 \\ \hline 
4 & (0.21,0.42)& 356    & 0.022601 & 0.114102 & 69.804714 & 2.396605 & 0.973349 &  0.086532 &  3778.7839 \\ \hline 
5 & (0.40,0.20)& 306    & 0.022609 & 0.113243 & 70.223014 & 2.334731 & 0.973016 &  0.074727 &  3779.3774 \\ \hline 
6 & (0.30,0.40)& 364    & 0.022601 & 0.114110 & 69.807095 & 2.399997 & 0.973447 &  0.087404 &  {\bf 3778.7830} \\  \hline 
7 & (0.35,0.25)& 374    & 0.022588 & 0.113895 & 69.868690 & 2.402573 & 0.972920 &  0.087475 &  3778.7909 \\  \hline
8 & (0.50,0.20)& 222    &0.022345  & 0.115229 & 68.930597 & 2.412138 & 0.965558 &  0.077042 &  3779.1554   \\  \hline
\end{tabular}
\caption{Best fit cosmological parameters for eight different initial conditions for Downhill-Simplex methods. Apart from 
the values of parameter $a$ and $b$ we have kept other parameters fixed i.e., $\epsilon=10^{-8}$ and $\alpha=-1,\beta=2$
and $\gamma=1/2$. How rapidly the algorithm converges depend on the values of the parameters $a$ and $b$ i.e., 
the initial simplex. The highest value of the likelihood function is shown in bold.}
\label{tab:amoeba1}
\end{center}
\end{table*}

Since the downhill simplex algorithm strictly demands that we should be able to compute
the (cost) function  at all the vertices at every step  therefor we must make sure that 
the values of the parameters always remain in the permitted range. We use the same range 
of cosmological parameters in {\tt COSMODSM}  which we use for {\tt COSMOPSO} 
and is given in Tab.~(\ref{tab:prior}).
Since there is no standard prescription for choosing the starting 
simplex in Downhill-Simplex method therefore we chose an ad-hoc prescription which is as
follows.

We represent the $i^{th}$ vertex of the simplex at iteration (or ``time'') $t$ by 
$X^I(t)$ and the minimum and maximum value of it are given by $X_{\rm min}$ and
$X_{\rm max}$. Note that here  ${X}$ is a  vector which has the dimensions of the 
parameter space (in our case it is six for the standard cosmological model). 

The first vertex of the initial simplex is set as :
\begin{equation}
X^1(0) = X_{\rm min} + a (X_{\rm max}-X_{\rm min}),  
\end{equation}
where $a$ is a user defined parameter in the range $[0,1]$ (we recommend it
to be less than 0.5). 

The other remaining $n$ vertices of the initial simplex are set as:
\begin{equation}
X^j(0) = X^1(0) +  \lambda_j  e^j, 
\end{equation}
where $j=2,.., n+1$ and $e^{j}$ are $n$ unit vectors. The coefficients  $\lambda_j$  
depend on the search range:
\begin{equation}
\lambda_j = b (X_{\rm max}^j-X_{\rm min}^j),
\end{equation}
where $b$ is a user defined parameter in the range $[0,1]$ (we recommend it
to be less than 0.5) and $j=1,...,n$.  Note that there is some degeneracy between 
$a$ and $b$  and in any ideal parametrization this should be avoided. Our parameterization 
is as good as choosing the initial points by trial and error. 

In {\tt COSMODSM}  at every iteration we find the vertices which have the largest and
smallest values of the cost function, called these $i_h$ and $i_l$ 
respectively. For better readability we also represent $X_{i_h}$ by $X_h$ and
$X_{i_l}$ by $X_l$.  The second highest point is also needed and we represent that by $X_{nh}$. 
We also compute the centeroid of $n$ vertices (keeping the $i_h$ aside)
and represent that with ${\bar X}$ with respect to which all the geometric
operations, reflection, expansion and contraction are carried out. 
We represent the value of the cost function by $f(X)$ and so :
\begin{equation}
f(X_h) \ge f(X_i) ~~~\text{for}~~~i=1,2,..,n+1,
\end{equation}
and
\begin{equation}
f(X_l) \le f(X_i) ~~~\text{for}~~~i=1,2,..,n+1.
\end{equation}

The centeroid ${\bar X}$ is defined as:
\begin{equation}
{\bar X} = \frac{1}{n} \sum\limits_{i=1, i\ne i_h}^{i=n+1} X^i 
\end{equation}

We use the same convergence criteria which is used in  \cite{1992nrfa.book.....P}
for which we compute:
\begin{equation}
\epsilon = 2 \frac{f(X_h)-f(X_l)}{|f(X_h)|+|f(X_l)|},
\end{equation}
and stop the search when $ \epsilon < \epsilon_0$ where $\epsilon_0$ is a user
defined number. 

In \cite{neldermean1965} three different equations are used for reflection, expansion 
and contraction, however, we use a single function $g(X,\xi)$  as defined below to carry out 
all the three operations:
\begin{equation}
g:X \longrightarrow X' = {\bar X} - \xi ({\bar X} - X), 
\end{equation}
where $ \xi=\alpha, \beta$ or $\gamma$ with $\alpha=-1, \beta > 1$ and $\gamma < 1$.
We consider $\beta=2 $ and $\gamma =1/2$ in the present work. 

The meaning of $\xi  $ can be easily understood by writing:
\begin{equation}
\xi = \frac{X'-{\bar X}}{X-{\bar X}},
\end{equation}
which means that $\xi$ is the ratio of the distance of the new point $X'$ and 
the old point $X$ from the centeroid ${\bar X}$.  

The main steps of the Nelder-Mead Algorithm can be summarized in the following way:

\begin{enumerate}
\item
Set up the initial simplex $X^i(0),i=1,2,..,n+1$, compute the values of the function 
$f_i(0)=f(X^i(0))$, and find out the highest (worst) and lowest (best)  point $X_h(0)$ and $X_l(0)$,
and the centeroid ${\bar X} $ of the best $n$ points. We also need to find the 
second worst (highest) point $X_{nh}$.  
\item
If the convergence is not reached reflect the worst point $X_h$ about the centeroid:
\begin{equation}
X_r = {\bar X} - \alpha ({\bar X} - X_h),
\end{equation}
and go to the next step. 
\item 
If we find $ f_l \le  f_r < f_{nh} $ than that means we have found a point which is better than 
the worst point $X_h$ so we replace $X_h$ with $X_r$ and recompute $X_l,X_h$ and ${\bar X}$ and
go to step (2). 
\item
If we find $ f_r  < f_l$ that means we are going in the right direction and it is useful
to expand the new point $X_r$ along the same direction and get a new point $X_e$:
\begin{equation}
X_e = {\bar X} - \beta ({\bar X} - X_r)
\end{equation} 
If we find that $f_e < f_r$ than we replace the worst point $X_H$ with $X_e$ otherwise we replace 
the worst point with $X_r$ and go to step (2). 
\item
If we find $ f_r \ge f_{nh}$ that means the simplex  is too big and we need contraction. However, before that we
test:
\begin{enumerate}
\item
If $ f_r \ge f_h$ i.e., the reflected point is worse than the worst or is as worse as the worst we contract 
\begin{equation}
X_c = {\bar X} - \gamma ({\bar X} - X_h).
\end{equation}
If we find that the point $X_c$ is worse than the worst $X_h$ we shrink the whole simplex around the 
best point $X_l$:
\begin{equation}
{\tilde X}_i = \frac{1}{2} (X_i+X_l),
\end{equation}
with $i=1,2,....,n+1$ and go to step (2) 
\item
If  $ f_r < f_h$ then we contract the reflected point $X_r$:
\begin{equation}
X_c = {\bar X} - \gamma ({\bar X} - X_r),
\end{equation}
and if  $f_c \le f_r$ then we replace $X_h$ with $x_c$  otherwise again shrink the simplex 
around the best point $X_l$ and go to step (2).
\end{enumerate}
\end{enumerate}

\begin{figure}
\begin{center}
\epsfig{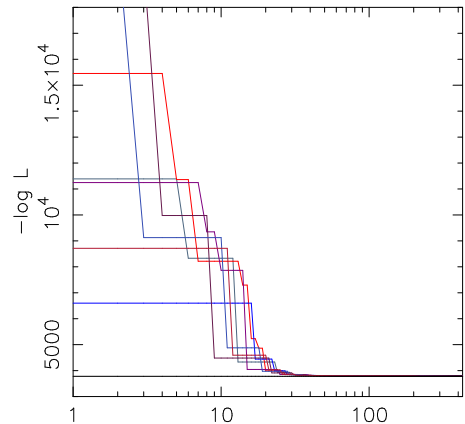}
\caption{Change in the cost (likelihood) function for the seven vertices we have (shown by different
colors) as the  Downhill Simplex algorithm iterates (x-axis shows the iterations) for the WMAP 
nine year data in {\tt COSMODSM}. The minimum (best) value of the  cost function is also shown by the straight line.}
\label{fig:amoeba_best_like}
\end{center}
\end{figure}

\begin{figure*} 
\begin{center}
\epsfig{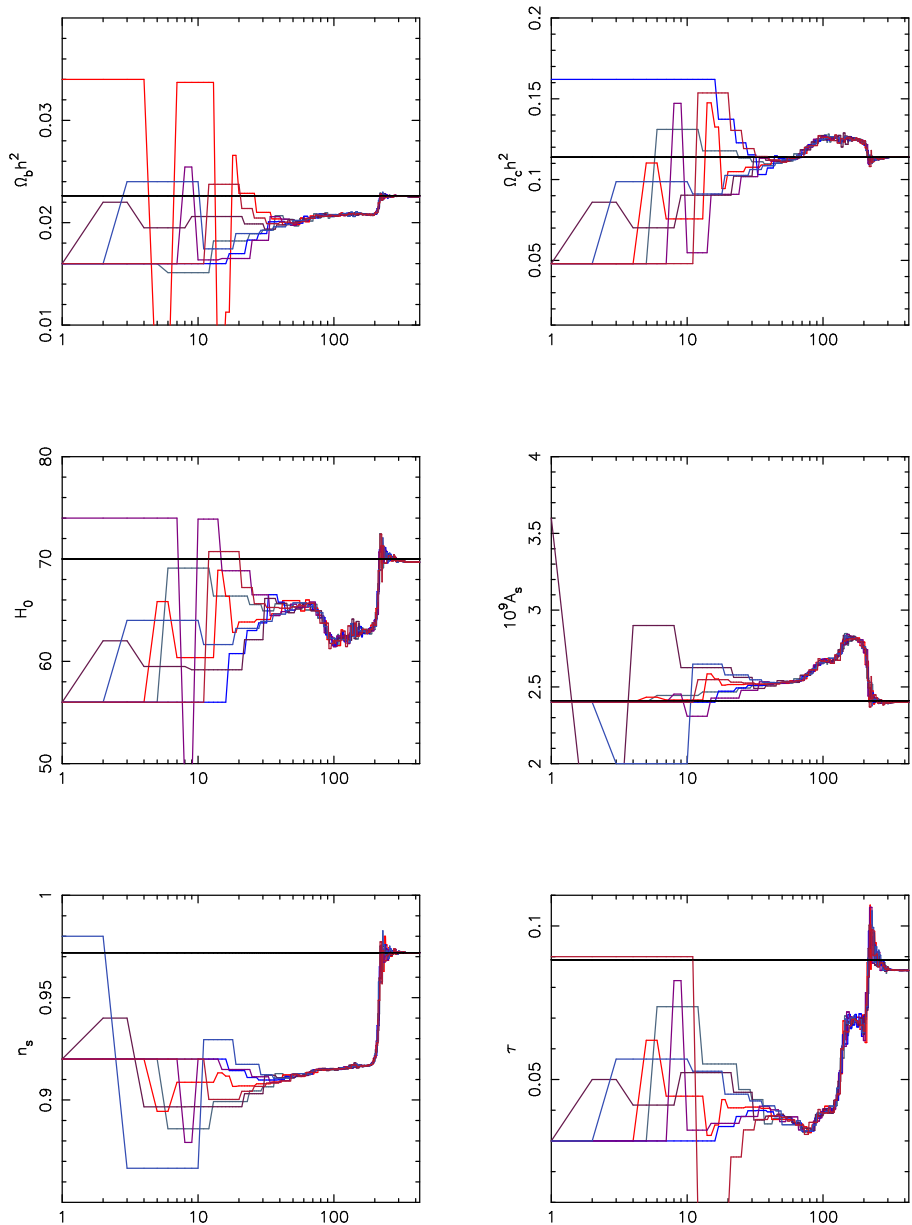}
\caption{Change in the vertices (shown by different colors) of the simplex with iterations 
(x-axis) along different directions i.e., cosmological parameters.
The best fit values of the parameters are  also shown by straight lines.}
\label{fig:amoeba_best}
\end{center}
\end{figure*}

\begin{figure*} 
\begin{center}
\epsfig{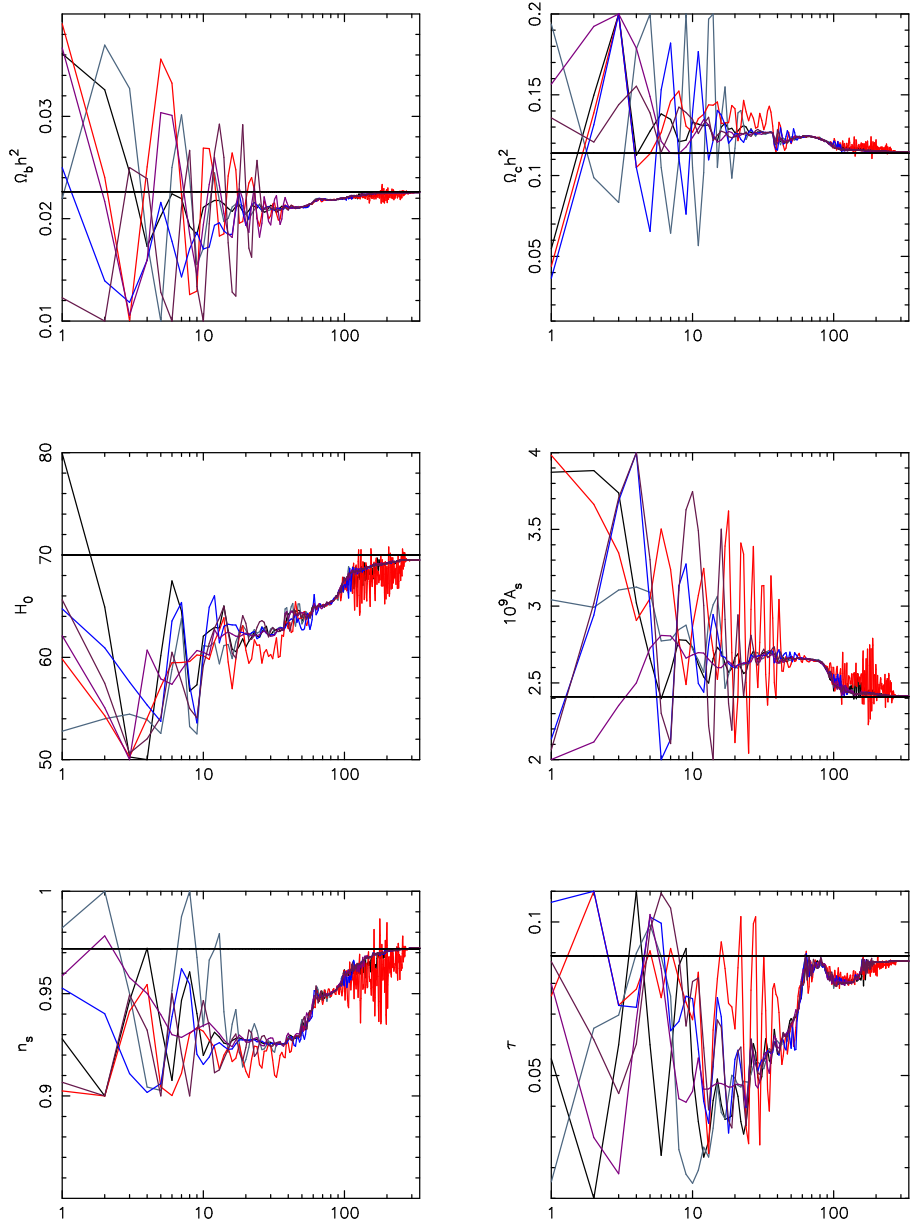}
\caption{Change in the positions of a few PSO particles with iterations (x-axis) 
along different directions. 
PSO particles sample much broader area of the parameters space along their journey towards 
the best fit point as compared to the vertices of simplex in Downhill Simplex method as shown in 
Fig.~(\ref{fig:amoeba_best}).}
\label{fig:pso_best}
\end{center}
\end{figure*}

In any iterative algorithm like PSO and downhill simplex in general it is not 
that important what convergence criteria we choose since most of them  
look for the progress made in each iteration and stop the search when the improvement 
falls below a tolerance limit or remains below a tolerance limit. 
All the three methods which discussed in the present work have different 
criteria, however, they all have a very good agreement with the final results. 

In order to demonstrate that the downhill simplex method of Nelder and Mead
can successfully find the best fit cosmological parameter we considered WMAP nine 
year data for our analysis. Apart from the programs for the exploration of parameter space,
we used the same program to compute the cost function (likelihood) which we use
in {\tt COSMOPSO} so there is no need to further elaborate on that here 
and so we will only present the results here. 

In Fig.~(\ref{fig:amoeba_best_like}) we show the change in the values of cost function 
at the vertices of the simplex as the simplex evolves. From the figure we can see that 
like in any other iterative method in DSM also in the beginning we have huge improvement in 
the cost function which falls rapidly which means in DSM also (like PSO) 
the convergence is reached only  asymptotically. 

We show the best fit cosmological parameters which we get for the WMAP nine year data 
using {\tt COSMODSM} in Tab.~(\ref{tab:amoeba1}) for eight different  
initial conditions i.e, different values of the parameters $a$ and $b$. 
We find that apart from  having different number of iterations for convergence they all
agree with the values of the best fit cosmological parameters with a reasonable limit. 

We also tried to see if DSM also can sample the parameter space like PSO 
and found that it does a very poor job and also suffer from the drawback of early convergence. 
In order to show that in Fig.~(\ref{fig:amoeba_best}) and (\ref{fig:pso_best}) we show the 
the vertices of simplex in {\tt COSMODSM} and trajectories of a few PSO particles
and {\tt COSMOPSO}  respectively. From these figures it is quite clear that PSO does 
sample a large volume of parameter space due to its  stochastic nature which is not true for DSM.  
 

\section{Powell's method of Bound Optimization BY Quadratic Approximation (BOBYQA)}

Since M. J. D. Powell gave his method of unconstrained optimization by quadratic
approximation in 1962 \cite{powel1962} there  have been  many changes 
in  his approach of non-linear optimization without derivative as is reflected 
by the software  {\tt COBYLA, NEWUOA, BOBYQA} and  {\tt LINCOA}  \cite{bobyqa} released by 
him time to time. In the present work we use {\tt BOBYQA}  FORTRAN 77 package which is 
available  for downloading from \cite{bobyqa}. 

Powell's method of Bound Optimization BY Quadratic Approximation (BOBYQA) is a technique
to find the global maximum/minimum of a function $f(X)$, $X \in {\mathcal R}^n$, with 
$ X_{\rm min} \le X \le X_{\rm max}$ without computing derivative of the cost function $f(X)$.
The method is based on approximating the function $f(X)$ by a quadratic function $q(X)$ at a 
set of  $m$ points $Y$ which are chosen and adjusted automatically. 
\begin{equation}
q(X) =  A + B^T(X-{\tilde X})  +  (X-{\tilde X})^T C (X-{\tilde X}), 
\end{equation}
where $A$ is a constant, $B$ a column vector with $n$ elements and $C$ is a $n\times  n$ symmetric 
matrix and in total we have  $m=1+n+n(n+1)/2=(n+1)(n+2)/2$ coefficients to fit the  quadratic function
for  which we consider a set of $m$ interpolating points $Y$ at  iteration $k$  at which the 
approximation holds: 
\begin{equation}
f_k(Y_j)  =  q(Y_j). 
\end{equation}
At every iteration $k$ we select a point $X_k$ from $Y_j$ which has the property:
\begin{equation}
f(X_k) = \text{min}(f(Y_j),j=1,...,m) 
\end{equation}
which we keep updating just like we keep updating the worst vertex at every iteration as we do in the 
Downhill-Simplex method. 

The quadratic approximation holds true only within a small region  characterized by a positive number $\Delta$ 
called the ``trust region radius'' for which we must give an initial $\Delta_{\rm beg}$  and final $\Delta_{\rm final}$ 
values. The value of $\Delta$ keep changing with iterations and once it falls below $\Delta_{\rm final}$ we stop
the program. Quadratic approximation within a small region is a common technique and is used
in many other algorithms also including one we published in \cite{2013PhRvD..88b3522G}. 
 
At iteration $k$ we construct a vector $d_k$ from $X_k$ such that, (1) $||d_k|| < \Delta_k$, 
(2) $X_{\rm min} \le  X_k+d_k \le X_{\rm max}$,
(3) $X_K+d_k$ is not one of $Y_j$.  Once we have a valid $d_k$ we can update $X_k$.
\begin{equation}
X_{k+1} = X_k + d_k, 
\end{equation}
if $ f(X_k+d_k) < F(X_k)$ otherwise we set $X_{k+1}=X_k$ and one of the interpolation points $Y_t$ is 
replaced with $X_k+d_k$. After this has been done we generate $q_{k+1}(X)$ and $\Delta_{k+1}$ for $k+1$ iteration 
according to the  prescription given in  \cite{bobyqa}. 

In order to use {\tt BOBYQA} for cosmological parameters estimation from the WMAP nine year data 
we made the following considerations. We compute theoretical CMB angular power spectrum $C_l$S using 
the publicly available code  {\tt CAMB} and use WMAP nine year likelihood code for computing the 
likelihood for temperature and polarization anisotropies as we do in section (II).  

Since  {\tt BOBYQA} is designed in such  a way that all the parameters must have values of the same 
order for which we normalize the six cosmological parameters with some default values so that they 
have the same order of $X_{\rm min}^i, X_{\rm max}^i$ and the same values of $X^i$ (where $i=1,...,6$) 
at every iteration. This step is important  otherwise the parameters like $H_0$ which has a large 
value will hardly get updated. 
We use default values $\Omega_bh^2=0.018,\Omega_ch^2=0.1,H_0=65.0,10^9A_s=2.2,n_s=0.94,\tau=0.08$ 
and avoided to give the exact values of the best fit cosmological parameters known to us. 

There is no prescription available for choosing the number of interpolation points $m$ and
one must find the appropriate value by trial and error. We have found that $m=n+10$, which is in the 
range $ [(n+2), (n+1)(n+2)/2]$, works fine for our case. For $\Delta_{\rm beg}$ and $\Delta_{\rm final}$ also 
there is no standard prescription so we must set their values by trying out different values. We use 
$\Delta_{\rm beg} = 10^{-2}$ and $\Delta_{\rm final}=10^{-12}$ and found that these values 
give  us acceptable convergence i.e., convergence is reached after 340 iterations. 

Here we would like to mention that all the versions of {\tt COSMOMC} \cite{cosmomc}
October 2012 onward also  use  Powell's {\tt BOBYQA} with some  modifications \cite{cosmomc} 
to find the best-fit cosmological parameters when we need only those. 
Our implementation of Powell's method  is completely independent and different
than what is used in {\tt COSMOMC} it terms of the uses and the values of design 
parameters $m$ and $\Delta_{\rm beg}$ and $\Delta_{\rm final}$. 
It is also important to keep in mind that MCMC methods themselves also can find the best-fit point 
without any external optimizer, however,  can take more time than {\tt COSMODSM } or {\tt COSMOBOBYQA}.   
\begin{table*}
\centering
\begin{tabular}{ccccc}\hline \hline 
Parameter       &  {\tt COSMOMC}           & {\tt COSMOPSO}   & {\tt COSMODSM}    & {\tt COSMOBOBYQA} \\ \hline  
$\Omega_bh^2$   & 0.022597       & 0.022601    &  0.022601   &  0.022630  \\ \hline 
$\Omega_ch^2$   & 0.112282       & 0.114044    &  0.114110   &  0.114148  \\ \hline
$H_0$           & 70.06592       & 69.82771    &  69.807095  &  69.87489  \\ \hline
$ 10^9A_s$      & 2.177424       & 2.395936    &  2.399997   &  2.414575  \\  \hline
$n_s$           & 0.973842       & 0.973390    &  0.973447   &  0.974537  \\ \hline
$\tau$          & 0.859690       & 0.086614    &  0.087404   &  0.091901  \\ \hline
$-\log {\mathcal L}$      & 3779.0220      & 3778.7839   &  3778.7830  &  3778.8495 \\ \hline  
\end{tabular} 
\caption{We show the best fit cosmological parameters, which give the maximum likelihood, 
for {\tt COSMOMC}, PSO, Downhill-Simplex methods and Powell's {\tt BOBYQA} in the second,
third, fourth and fifth columns of the above table respectively.}
\label{tab:all_methods}
\end{table*}

In Tab.~(\ref{tab:all_methods}) we show the cosmological parameters which we have estimated 
with four different methods named {\tt COSMOMC}, {\tt COSMOPSO}, {\tt COSMODSM} and  {\tt COSMOBOBYQA}
and found that all the methods are able to find the correct values of the best fit cosmological parameters  
within reasonable time.

\section{Discussion and Conclusion}

Building on our earlier work \cite{2012PhRvD..85l3008P} in the present work we have shown that
particle swarm optimization not only can be used to find the best fit cosmological
parameters from CMB data sets like WMAP and Planck, it can also be used to sample the 
parameter space quite effectively. We have found that when using PSO as a sampler we must
avoid early convergence and use multiple realizations of PSO  so that the parameter 
space is covered uniformly. We present the results of our analysis of PSO for the WMAP nine 
year and Planck data and show that these are quite consistent with the standard results. 
In order to show that the sampling of the parameter space is quite effective we show that
we can use the same analysis pipeline {\tt GetDist} to process the PSO sampled point which is used
to process the MCMC sampled points in {\tt COSMOMC} and get consistent results which 
is clear from the one dimensional marginalized probability distribution and 68\% and 95\% 
confidence regions we show in Figs.~(\ref{fig:pso_cont_w9}) and ({\ref{fig:pso_cont_w9p}) 
for the WMAP nine year and WMAP nine year + Planck data respectively. 

 We ran {\tt COSMOMC} and {\tt COSMOPSO}   
on a Linux cluster with multiple nodes using Message Passing Interface (MPI) library and  
shared memory parallelization library {\tt OpenMP}.  We have found that {\tt COSMODSM} 
and {\tt COSMOBOBYQA} are faster than {\tt COSMOPSO} and {\tt COSMOMC},  however, need 
fine tuning of the design parameters  and have higher chances of failure. {\tt COSMOPSO}
has the problem of early convergence but that is not an issue if our aim is just to find
the best fit point and not to sample the parameter space. In general, for 30 PSO particles
in {\tt COSMOPSO} we can achieve convergence within 400 steps, or less than 12000 computation 
of the  cost function ({\tt CAMB} calls + likelihood computation). For {\tt COSMODSM} we
need 400 steps with 7 vertices or around 2800 computation of the cost function. 
For {\tt COSMOBOBYQA}  we  achieve convergence in around 400 steps or 400 
computation of the cost function.   It should be noted that all the different methods
we have discussed here have different convergence criteria so direct comparison is
not fair, however, whatever we have mentioned is qualitatively is true i.e., {\tt COSMOBOBYQA} 
method is faster than {\tt COSMOPSO}, {\tt COSMODSM} and {\tt COSMOMC}.

In the present work we  use  only  basic formalism of PSO, however, PSO
has scope for modifications also some of  which have been already  done for different type of 
problems (see \cite{Blum2008,Lazinic2009}). We believe that with some modifications PSO 
should be able to become a robust method for sampling also in the Bayesian data analysis. 
One of the interesting properties of PSO is that the trajectories of PSO 
particles can be compared with trajectories of point particles in classical 
mechanics  since they also follow trajectories which minimize some function 
(action) like PSO particles. However, this analogy cannot be stretched too 
far since for PSO particles time is not continuous i.e.,  $ t \in \mathbb{Z}$.

We also presented implementations of two new methods, Downhill simplex method of 
Nelder and Mead and Powell's method of  Bound Optimization BY Quadratic Approximation
(BOBYQA) in the present work which are must faster than PSO but are less robust, in 
the sense, PSO is very less sensitive to its design parameters and does not need
any starting point (all the starting points are equally good) which is not the case
for other algorithm.  Since {\tt COSMOPSO} is a completely parallel code
(uses  distributed memory parallelism using MPI and shared memory parallelism 
using OpenMP) therefore computational cost is not an issue for a cluster. 

We have found that all the four methods (as are summarized in  Tab.~(\ref{tab:all_methods})
are quite successful in finding the best-fit cosmological parameters in a reasonable time. 

Cosmological parameter estimation from CMB data sets is still an  active area of
research in cosmology and optimization methods like PSO can be quite helpful
not only in cosmological parameters but at other stages also of CMB data analysis
pipeline. Being an independent and robust PSO has potential of becoming an
essential component of the CMB data analysis toolkit.

\vspace{1.0cm}  
\centerline{\bf Acknowledgments}
\vspace{1.0cm}  

The author acknowledges  the use of codes {\tt CAMB}, {\tt COSMOMC} and Planck likelihood {\tt Plc}  
for this work and also thank  the  WMAP and Planck team for making the data publicly available.
The authors thanks Scott Dodelson for suggesting to compare PSO with the downhill simplex method
and Gaurav Goswami for the feedback and comments on the drafts. The author would also like
to thank Tarun Souradeep for giving insight about CMB physics time to time
and introducing the problem of cosmological parameter estimation. 
All the computational work for this project was done on the IUCAA HPC facility.  
The author specially thank the Science and Engineering Research Board (SERB) 
of the Govt. of India for financially supporting this work via 
a Start-Up Research Grant (Young Scientists) with grant no SR/FTP/PS-102/2012.

\bibliographystyle{h-physrev3}
\bibliography{cmbr}

\end{document}